\documentclass[11pt]{article}

        \usepackage{setspace,fullpage,epsfig,graphics,amsbsy,amssymb,cancel,slashed,mathrsfs,booktabs,physics}
    \usepackage{psfrag,hyperref}
    \usepackage{graphicx,color}
    \usepackage{wrapfig}
    \usepackage{subfigure}
    \usepackage{booktabs}
    \usepackage{titlesec}
\hypersetup{colorlinks=true}
\hypersetup{linkcolor=black}
\hypersetup{citecolor=red}
\hypersetup{urlcolor=blue}
    \usepackage[nameinlink]{cleveref}

    \newcommand{\be}{\begin{equation}}
    \newcommand{\ee}{\end{equation}}
    \newcommand{\bes}{\begin{split}}
    \newcommand{\ees}{\end{split}}
    \usepackage{amsmath}
    \numberwithin{equation}{section}
    \usepackage{caption}
    \usepackage[nosort]{cite}

    \def\al{\alpha}
    \def\eps{\epsilon}
    \newcommand{\bea}{\begin{eqnarray}}
    \newcommand{\eea}{\end{eqnarray}}  
    \newcommand{\nn}{\nonumber}

    \newcommand{\NN}{\mathcal{N}}
    \newcommand{\sfrac}[2]{\mbox{$\frac{#1}{#2}$}}

    \newcommand{\LL}{{\mathcal L}}

    \def\bS{\mathbb{S}}
    \def\s{\sigma}

    \def\bR{\mathbb{R}}
    \newcommand{\sbkt}[1]{\left[#1\right]}
    \newcommand{\bkt}[1]{\left(#1\right)}
    \newcommand{\p}{\partial}

    \usepackage{amsmath}

    \usepackage[left=2.5cm,right=2.5cm,top=2.5cm,bottom=3cm]{geometry}
\titleformat{\section}{\normalfont\bfseries}{\thesection.}{4pt}{}
\titlespacing{\section}{0pt}{20pt}{6pt}

\titleformat{\subsection}{\normalfont\bfseries}{\thesubsection.}{4pt}{}
\titlespacing{\subsection}{0pt}{15pt}{6pt}

\titleformat{\subsubsection}{\normalfont\itshape}{\thesubsubsection.}{4pt}{}
\titlespacing{\subsubsection}{0pt}{15pt}{6pt}

\DeclareFontShape{OT1}{cmr}{mx}{n}%
    {<->cmr10}{}
\newcommand{\mytitlefont}{\fontseries{mx}\selectfont}
\DeclareMathAlphabet{\titlemath}{OT1}{cmr}{mx}{n}

\begin{document}

% TITLEPAGE

\begin{titlepage}
\thispagestyle{empty}
\begin{flushright} \small
UUITP-52/18\\MIT-CTP/5084
 \end{flushright}
\smallskip
\begin{center}

~\\[2cm]

{\fontsize{26pt}{0pt} \mytitlefont Gauge theories on spheres with 16 supercharges and non-constant couplings}

~\\[0.5cm]

Joseph A. Minahan\,$^{1}$  and Usman Naseer\,$^{1,2}$

~\\[0.1cm]

$^1$~{\it Department of Physics and Astronomy,
     Uppsala University,
     Box 516,
     SE-751s 20 Uppsala,
     Sweden}

$^2$ {\it Center for Theoretical Physics,
     Massachusetts Institute of Technology,
     Cambridge, MA 02139, USA.}

~\\[0.8cm]
\end{center}
We construct a class of theories with 16 supersymmetries on spheres of dimension nine and less.
The gauge coupling and mass terms for the scalar fields depend on the polar angle away from the north pole. 
Assuming  finite coupling on the north pole, this leads to zero coupling at the south pole  for $d>4$ and infinite coupling at the south pole for  $d<4$. 
The underlying supersymmetry algebra of these theories is shown to be isomorphic to the Poincar\'e superalgebra in $d$-dimensions.  We also give a localization procedure which  leads to nontrivial results for $d=2$. 

\noindent  

\vfill

\begin{flushleft}
%\today
\end{flushleft}

\end{titlepage}

% TABLE OF CONTENTS

\tableofcontents

%%%%%%%%%%%%%%%%%%%%%%%%%%%%%%%%%%%%%%%%%%%%%%%%
\section{Introduction}
The study of supersymmetric  gauge theories on compact manifolds has led to many  advances 
  beyond perturbation theory.
 Several exact results for the partition functions, indices, Wilson loops and other supersymmetric observables have now been obtained, providing highly non-trivial checks\footnote{For reviews see \cite{Pestun:2016zxk} and references therein.}.
The calculations for these observables rely on  localization, which in turn depends on preserving some of the global supersymmetries on the compact manifold. 
 The size of the  manifold serves as a natural regulator for the IR divergences. 
 
 In general it is not possible to put a supersymmetric gauge theory on a curved manifold without breaking all of the supersymmetries.
 If the manifold admits covariantly constant spinors then one can preserve at least some of the supersymmetries by replacing all derivatives that appear  in the Lagrangian and supersymmetry transformations with covariant derivatives. If no such spinors exist then this \emph{minimal coupling} will not be supersymmetric on the curved space. However, depending on the manifold, it may be possible to add additional  terms to the Lagrangian and supersymmetry transformations to preserve the supersymmetry.   One systematic approach \cite{Festuccia:2011ws} is  to consider an off-shell supergravity theory coupled to matter multiplets. Then one can give background values to the gravity multiplet that preserve supersymmetry and take the Planck mass to infinity. This decouples the gravitational dynamics and leaves  a global supersymmetric theory on the curved space. However, this approach is limited to examples where one knows an off-shell formulation of supergravity. 
  
One of the simplest curved spaces which  admits global supersymmetries is a $d$-dimensional sphere, $\bS^d$. Since $\bS^d$ is conformally flat there is a canonical way to put a superconformal field theory (SCFT) on $\bS^d$. 
In \cite{Pestun:2007rz} Pestun studied supersymmetric gauge theories on $\bS^4$ with eight supercharges. Following his seminal work,  
 supersymmetric theories on spheres of dimension $d\leq 7$ with different number of supercharges have been studied in \cite{Kapustin:2009kz, Kim:2012ava,Kallen:2011ny,Kallen:2012cs,Kallen:2012va,Minahan:2015jta}. A uniform approach for perturbative supersymmetric gauge theories on $\bS^d$ was  given  in \cite{Minahan:2015any,Gorantis:2017vzz}. 
  
 Supersymmetric gauge theories with 16 supercharges and a constant coupling can only be placed on spheres with $d\le7$, and with 8 supercharges only  on spheres with $d\le5$.  One can understand this by considering the available supergroups where the supercharges transform in spinor representations.  For example, on $\bS^7$ there exists the group $OSp(8|2)$ which has a bosonic subgroup $SO(8)\times SU(2)$ corresponding to the isometry of the sphere and the $R$-symmetry.  Likewise, on $\bS^5$ one can have the supergroup $SU(4|1)$ with 8 supersymmetries and the bosonic symmetry group $SU(4)\times U(1)$.  An implicit assumption in this argument is that the bosonic subgroup contains the full isometry group of the sphere. However, it is possible that the Lagrangian explicitly breaks part of the sphere isometry group. 
Such constructions, though somewhat exotic, are not unfamiliar from the perspective of field theory, e.g., see \cite{Gaiotto:2008sd,Choi:2017kxf,Maxfield:2016lok,Festuccia:2016gul,Goto:2018bci} for supersymmetric theories with varying coupling, theta angles and K\"ahler moduli. More recently (1,0) supersymmetric theories on $\bS^6$ were constructed with a non-constant coupling \cite{Naseer:2018cpj}.

In this paper we explore this possibility more generally and construct a  class of supersymmetric theories on $\bS^d$ where the coupling is not constant on the sphere, but depends on the polar angle. In this construction, if $d>4$  the gauge coupling is zero at the south pole and smoothly varies to a non-zero value at the north pole. If $d<4$ the gauge coupling is infinite at the south pole.  The Lagrangian is invariant only under an $SO(d)$ subgroup of the  isometry group $SO(d+1)$ of $\bS^d$. Hence the no-go arguments stated  above do not  directly apply. 

Our construction can be applied to gauge theories with 16 supersymmetries on $\bS^d$, with  $d\leq9$. To construct these theories, we dimensionally reduce 10D $\NN=1$ SYM onto $\bR^{1,9-d}\times \bS^d$. We then allow for the gauge coupling and the mass term for the scalars to be non-constant. A careful analysis of the supersymmetry parameters then gives their position dependence. We also give a formulation of our construction with one off-shell supersymmetry. 
For $d= 8,9$  this modification of the coupling circumvents the previous restriction for putting the theory on the sphere.
 For $d\leq 7$ this gives new supersymmetric theories with same field content as that of the maximally supersymmetric theories.
  We also discuss the symmetry algebras for these theories. We show that the underlying superalgebra is not semisimple and  is \emph{isomorphic} to the Poincar\'e super algebra found in flat space. 
 
This last fact is not surprising since the gauge coupling we find is essentially a conformal compensating factor for the theory on the sphere.    
In other words, the angular dependence of the coupling compensates for the conformal transformation mapping the flat space theory with constant coupling to the sphere.
We emphasize however that the theories we consider are compact; the point at infinity, namely the south pole, is included.  In fact the same phenomenon is present for maximally supersymmetric gauge theories on $\bS^4$.  Pestun showed that the inclusion of the point at infinity leads to the inclusion of instantons localized at the south pole which contribute in a nontrivial way to the partition function \cite{Pestun:2007rz}.  While we do not compute partition function in this paper, we expect a similar phenomenon to occur.  

We then show how to localize these theories.  In  Pestun's construction the theory is localized by choosing a  a linear combination of spinors associated with a super Poincar\'e and a superconformal  transformation and then adding to the Lagrangian a term that is $Q$-closed under this linear combination.  Here we localize with a super Poincar\'e spinor only.   With this choice we then construct a $Q$-closed term to add to the Lagrangian which is analogous to the one in the Pestun construction.  Here we will find that the fields localize to commuting scalar fields that are
 solutions of the equations of motion on sphere.

At the localization locus the Yang-Mills action is divergent for nonzero scalar fields if $d>2$ and zero if $d<2$.  However, for $d=2$ one finds a finite action, suggesting that in this case one can proceed with this localization procedure  to extract nontrivial behavior for the free energy and BPS Wilson loops.  At the localization locus the auxiliary fields are set to zero, hence the localization is on-shell.  Naively, this would lead to a zero action.  But because of the compactness of the manifold, the action has a contribution at infinity that gives a nontrivial result.
We then consider a different $Q$-closed term to add to the action, which is reminiscent of the Higgs branch localization construction in \cite{Benini:2012ui}.

This paper is organized is as follows: In \cref{sec:review} we review the procedure of dimensional reduction of 10D $\NN=1$ SYM to obtain supersymmetric theories on $\bS^d$. In \cref{sec:GenD}, we construct theories with sixteen supersymmetries and non-constant gauge coupling on $\bS^d$ for $d\leq 9$. In \cref{sec:salgebra} we determine the supersymmetry algebra of these theories and compare it with their flat space counter parts. In \cref{sec:localization} we comment on the localization of path integral for these theories. In \cref{sec:conc} we present our conclusions and discuss further issues. The appendices contain our conventions and technical details of various computations.
\section{MSYM via dimensional reduction: Review}
\label{sec:review}
In this section we review the procedure in  \cite{Minahan:2015jta} to construct supersymmetric gauge theories on $\bS^d$. This is a generalization of Pestun's study in four dimensions \cite{Pestun:2007rz}. 
Our starting point is the 10 dimensional $\NN=1$ SYM Lagrangian\footnote{As in \cite{Pestun:2007rz} we consider the real form of the gauge group so that the group generators are anti-Hermitian and independent generators satisfy $\Tr(T^aT^b)=-\delta^{ab}$.}
\begin{equation}\label{LL}
\LL= -\frac{1}{g_{10}^2}\Tr\left(\sfrac12F_{MN}F^{MN}-\Psi\slashed{D}\Psi\right)\,,
\end{equation}
The space-time indices  $M,N$  run from $0$ to $9$ and $\Psi^a$ is a Majorana-Weyl spinor in the adjoint representation.  Properties of  $\Gamma^{M}_{ab}$ and $\tilde\Gamma^{M\,ab}$ are given in \cref{conv}.  The 16 independent supersymmetry transformations that leave \cref{LL} invariant are
\begin{equation}\label{susy}
\begin{split}
\delta_\eps A_M&=\eps\,\Gamma_M\Psi\,,\\
\delta_\eps \Psi&=\sfrac12 \Gamma^{MN}F_{MN}\,\eps\,,
\end{split}
\end{equation}
where $\eps$ is a constant bosonic  real spinor, but is otherwise arbitrary.

We next dimensionally reduce this theory to $d$ dimensions by choosing Euclidean spatial indices $\mu=1,\dots d$ with gauge fields $A_\mu$ and scalars $\phi_I$ with $I=0,d+1,\dots 9$.  The field strengths with  scalar indices become $F_{\mu I}=D_\mu \phi_I$ and $F_{I J}=[\phi_I,\phi_J]$.  As in \cite{Pestun:2016zxk} we choose one scalar to come from dimensionally reducing the time direction, leading to a wrong-sign kinetic term for this field.

We take the  $d$-dimensional Euclidean space to be the round sphere $\bS^d$ with radius $r$ and  metric
\begin{equation}\label{metric}
ds^2=\frac{1}{(1+\beta^2x^2)^2}\, dx_\mu dx^\mu\,,
\end{equation}
where $\beta=\tfrac{1}{2r}$. 
The supersymmetry parameters are modified to be conformal Killing spinors  (CKS) on the sphere, satisfying the conformal Killing spinor equations (CKSE)
\begin{equation}\label{KS1}
\nabla_\mu\eps=\tilde\Gamma_\mu\tilde\eps\,,\qquad\qquad \nabla_\mu\tilde\eps=-\beta^2\Gamma_\mu\eps\,.
\end{equation}
The general solution is
\be \label{eq:ckssol}
\eps = \frac{1}{\bkt{1+\beta^2 x^2}^{1/2}}\bkt{\eps_s+\beta x\cdot \tilde{\Gamma} \tilde\eps_c},
\ee
where $\eps_s$ and $\tilde\eps_c$ are constant but otherwise independent spinors. This solution has $32$ independent parameters. 
We impose the further condition
\begin{equation}\label{KS}
\nabla_\mu\eps=\beta\,\tilde\Gamma_\mu\Lambda\, \eps\,,
\end{equation}
leaving 16 independent supersymmetry transformations. To be consistent with \cref{KS1}, $\Lambda$ must satisfy $\tilde\Gamma^\mu\Lambda=-\tilde\Lambda\Gamma^\mu$,  $\tilde\Lambda\Lambda=1$, $\Lambda^T=-\Lambda$.  The simplest choice is $\Lambda=\Gamma^0\tilde\Gamma^8\Gamma^9$ which gives supersymmetric gauge theories on $\bS^d$ with $d\leq 7$.

On the sphere the supersymmetry transformations for the bosons are unchanged, but those for the  fermions are modified to
\begin{equation}\label{susysp1}
\begin{split}
\delta_\eps \Psi=&\sfrac12 \Gamma^{MN}F_{MN}\eps+\frac{\alpha_I}{2}\Gamma^{\mu I}\phi_I\nabla_\mu\,\eps\,,
\end{split}
\end{equation}
where the constants $\al_I$ are given by
\begin{equation}
\label{alrel}
\begin{split}
\alpha_I=&\frac{4(d-3)}{d}\,\quad \text{for}\quad I=8,9,0.\qquad 
\alpha_I=\frac{4}{d}\,\quad \text{for}\quad I=d+1,\dots 7\, .
\end{split}
\end{equation}
The index $I$ in \cref{susysp1} is summed over. This particular choice  preserves all 16 supercharges if one adds the following extra terms to the Lagrangian.
\begin{equation}\label{eq:extra}
\begin{split}
\LL_{\Psi \Psi} & =\ -\frac{1}{g_{\text{YM}}^2} \Tr \frac{(d-4)}{2r}\Psi\Lambda\Psi, \\
 \LL_{\phi\phi}&=- \frac{1}{g_{\text{YM}}^2}\left(\frac{d\,\Delta_I}{2\,r^2}\,\Tr \phi_I\phi^I\right)\,, \\ 
 \LL_{\phi \phi \phi }\ & = \frac{1}{g_{\text{YM}}^2}\frac{2}{3r}(d-4) 
 \varepsilon_{ABC} \Tr\bkt{  [\phi^A,\phi^B]\phi^C}.
\end{split}
\end{equation}
Here $\Delta_I$ is defined as 
\begin{equation}
\Delta_I= \alpha_I, \qquad \text{for}\qquad I=8,9,0, \qquad \Delta_I\ =\ 2\frac{d-2}{d}\qquad \text{for}\qquad I=d+1,\cdots 7.
\end{equation}
The scalars  split into two groups, $\phi^A$, $A=0,8,9$ and $\phi^i$, $i=d+1,\cdots 7$ and the $R$-symmetry is manifestly  broken from $SO(1,9-d)$ to 
$SO(1,7-d)$. 
The full supersymmetric Lagrangian is the dimensionally reduced version of~\cref{LL} supplemented with $\LL_{\Psi \Psi}, \LL_{\phi \phi}$ and $\LL_{\phi\phi\phi}$.

The number of supersymmetries can be reduced by imposing further consistent projections on the supersymmetry parameters. Field content then splits into the vector-multiplet and hypermultiplet or chiral-multiplet depending on the number of supersymmetries. For one particular Killing spinor, the supersymmetry can be realized off-shell by using pure spinors and seven auxiliary fields. The auxiliary fields also split into components of the vector multiplet, hypermultiplet and chiral multiplet.
 
 %%%%%%%%%%%%%%%%%%%%%%%%%%%%%%%%%%%%%%%%%%%%%%%%
\section{MSYM with non-constant coupling on $\bS^d$}
\label{sec:GenD}

%%%%%%%%%%%%%%%%%%%%%%%%%%%%%%%%%%%%%%%%%%%%%%%%

In this section we  give a different procedure for dimensional reduction which employs a non-constant coupling.
This procedure gives supersymmetric gauge theories on $\bS^d$ for $d\leq 9$ with sixteen supercharges. We also give an off-shell formulation for one of the supercharges. 
 \subsection{On-shell considerations}
 \label{s-onshell-susy}

 In contrast to the theories considered in the review, we now allow for a position dependent coupling constant and a position dependent mass term for the scalar fields on the sphere.   We  assume that the Lagrangian  in $d$ dimensions has the form
 \be\label{LL1}
\LL= \frac{e^{\phi}}{g_{\text{YM}}^2}\Tr\left(\sfrac12F_{MN}F^{MN}-\Psi\slashed{D}\Psi+\frac{d\,\Delta}{2\,r^2}\,\phi_I\phi^I\right)\,,
 \ee
 where $\phi$ and $\Delta$ may have position dependence on the sphere.
 
 For the case of constant coupling, the number of independent components of the conformal Killing spinor are reduced  to 16 by imposing \cref{KS}. This corresponds to choosing a particular set of 16 linear combinations of the components contained in $\eps_s$ and $\tilde{\eps}_c$ as the set of generators of the supersymmetry transformations. 
However, one could try imposing different conditions to reduce the supersymmetry.  
Here we propose the condition that $\tilde\eps_c=0$, such that $\tilde\eps=\partial_\mu f\Gamma^\mu\eps$, with $f$ given by
\be \label{eq:fdef}
f=-\frac{1}{2}\log(1+\beta^2x^2)\,.
\ee
The Lagrangian in \cref{LL1} is then invariant under the supersymmetry transformations 
 \be\label{susysp}
 \begin{split}
 \delta_\eps A_M=&\eps\,\Gamma_M\Psi\,,  \\
  \delta_\eps \Psi=&\sfrac12 \Gamma^{MN}F_{MN}\eps+\frac{2}{d}\,\Gamma^{\mu I}\phi_I\nabla_\mu\,\eps\,,
\end{split}
 \ee
 provided that $\phi$ and $\Delta$ have a certain profile on the sphere.

To verify this
claim, we start by computing the supersymmetry variation of the Lagrangian in \cref{susysp}.  This is done in detail in \cref{app:delLgenD}, with the result given in \cref{eq:delLgenD} which reads
\be\label{eq:varLLNonCons}
\begin{split}
\delta\LL=&\frac{e^\phi}{g_{\text{YM}}^2}\Big(-(d-4)F^{MN}\tilde\eps\Gamma_{MN}\Psi-4\phi^I(\nabla^\mu\tilde\eps)\Gamma_{I\mu}\Psi
+4d\beta^2\Delta\,\phi^I\eps\Gamma_I\Psi
 \\
&\qquad\frac{1}{2}\p_\nu \phi \left(F^{MN}\eps\Gamma^{MN}\Gamma^{\nu}\Psi-2F^{MN}\eps\Gamma^{MN\nu}\Psi-4\phi^I\tilde\eps\Gamma^I\Gamma^\nu\Psi\right) 
\Big).
\end{split}
\ee
Setting $\tilde\eps=\partial_\nu f\Gamma^\nu\eps$ and using  
\be
\Gamma^\nu\Gamma^{MN}=2\Gamma^{MN\nu}-\Gamma^{MN}\Gamma^\nu\,,
\ee
we see that the $F^{MN}$ terms cancel if
\be
 \partial_\nu \phi=-2(d-4)\partial_\nu f\,,
\ee
and hence
\be
e^\phi=(1+\beta^2x^2)^{d-4}= g^{\frac{4-d}{2d}},
\ee
where $g$ is the determinant of the metric on the sphere. We have dropped an overall constant which can be absorbed into $g_{\text{YM}}^2$.  
This gives an effective gauge coupling which varies over the sphere
\be
g_{\text{eff}}^2 = g_{\text{YM}}^2 \bkt{1+\beta^2 x^2}^{4-d}.
\ee
The coupling is finite at the north pole which corresponds to $|x|=0$, and flows to zero coupling at the south pole at $|x|\to \infty$ if $d>4$.  For $d=4$, the coupling stays constant. For $d<4$, the coupling becomes infinite at the south pole. 

Using the second equation in (\ref{KS1}) we find that cancellation of the $\phi^I$ terms requires setting

\be\label{Deltaeq}
\Delta=1-\frac{1}{2 d \beta^2} \p_\nu \phi \partial^\nu f=1+\frac{d-4}{d}\beta^2x^2 \, = 1+\frac{d-4}{ 4 d^3 }  r^2 g^{\mu\nu} \p_\mu \log g \p_\nu \log g\,.
\ee
This term blows up at the south pole, and the scalars become infinitely massive at the south pole. 
This, however, is an artifact of our choice of field variables. Namely, we can write the Lagrangian in terms of the ``canonical" fields and coupling by rescaling all  fields by a factor of $g_{\text{eff}}\bkt{x}$. A short computation then gives the Lagrangian 
\be
\label{eq:canLag}
\LL\, = \, \tfrac12 F_{\mu\nu} F^{\mu\nu} + \Psi \slashed{D} \Psi + D_{\mu}\phi_I  D^{\mu}\phi^{ I} +\frac{d\bkt{d-2}}{4 r^2}\phi^I \phi_I+ \tfrac12 g_{\text{eff}}^2\,  \sbkt{\phi_I,\phi_J}\sbkt{\phi^I,\phi^J},
\ee
where the covariant derivative is now $D_\mu\equiv\p_\mu + g_{\text{eff}}\,  A_\mu $. The mass term for the scalars is just the conformal  mass term on $\bS^d$. 

Notice that for $d=4$,  the coupling and  masses are not position dependent. In this case the dimensionally reduced Lagrangian is that of the $\NN=4$ SYM conformally coupled to the sphere.
Also note that we could have instead chosen $\eps_s=0$, and hence $\tilde\eps=\frac{1}{x^2}\, x\cdot\Gamma\eps$.  In this case 
\be
f=-\frac{1}{2}\log(1+1/(\beta^2 x^2))
,\quad 
e^\phi=(1+1/(\beta^2x^2))^{d-4} 
,\quad 
\Delta=1+\frac{(d-4)}{\beta^2x^2}
\ee
Here the behavior at the poles is reversed.

This gives a consistent construction of supersymmetric gauge theories with 16 supercharges on $\bS^d$ with $9\geq d \geq 4$. One can wonder if number of supercharges can be reduced in this construction to obtain theories with eight and  four supercharges  on $\bS^6
$ and $\bS^4$ respectively ---
cases where the previous construction did not work. However a close inspection of the possible ways to reduce the number of supercharges shows that the situation is not improved. 
A possibility is to impose the constraint $\Gamma^{09} \eps=\eps$. However one cannot consistently impose this constraint on the field content. Fermions with opposite eigenvalues of operator $\Gamma^{09}$ are coupled via their kinetic term.

%%%%%%%%%%%%%%%%%%%%%%%%%%%%%%%%%%%%%%%%%%%%%%%%
  \subsection{Off-shell supersymmetry}
  \label{s-offshell-susy}
%%%%%%%%%%%%%%%%%%%%%%%%%%%%%%%%%%%%%%%%%%%%%%%%
  To realize supersymmetry off-shell we proceed by choosing a set of pure spinors $\nu^m$ for $ m=1,2,\cdots,7$. They satisfy the completeness conditions given in \cref{psprop}.  
The off-shell supersymmetry transformations are 
 \be\label{susyos}
 \begin{split}
 \delta_\eps A_M=&\eps\,\Gamma_M\Psi\,,  \\
  \delta_\eps \Psi =&\sfrac12 \Gamma^{MN}F_{MN}\eps+\frac{2}{d}\Gamma^{\mu I}\phi_I\nabla_\mu\,\eps+K^m\nu_m\, ,  \\
\delta_\eps K^m=&-\nu^m\slashed{D}\Psi-\frac{1}{2}
\p_\mu \phi
\nu^m\Gamma^\mu\Psi\,.
\end{split}
 \ee
 It is straightforward to check that the Lagrangian is invariant under these transformations provided we include the additional term
 \be
 \LL_{\text{aux}}=-\frac{e^{\phi}}{g_{\text{YM}}^2}\,\Tr K^mK_m\,.
 \ee
 
 We then must show that the transformations in (\ref{susyos}) close up to a symmetry of the action.  To this end, we compute the action of two successive supersymmetry transformations on fields.  For the vector field $A_\mu$, we find
 \be
 \begin{split}
 \delta^2_\eps A_\mu=\delta_\eps(\eps\Gamma_\mu \Psi)=&\frac12F^{MN}\eps\Gamma_\mu \Gamma_{MN}\eps+{2}\phi_I\eps\Gamma_I \tilde\Gamma_\mu\tilde\eps+K^m\eps\Gamma_\mu\nu_m  \\
 =&\eps\Gamma^\nu\eps F_{\mu\nu}+\eps\Gamma_I \eps D_\mu\phi^I+2\phi^I\eps\Gamma_I\nabla_\mu\eps  \\
 =&-v^\mu \nabla_\mu A_\nu -A_\nu \nabla_\mu v^\nu + D_\mu \bkt{v^M A_M},
 \end{split}
 \ee
 where $v^M \equiv \eps \Gamma^M \eps$. The first two terms in $\delta_\eps^2$ form a Lie derivative and the last term is a gauge transformation.
 
For the scalar fields we have
 \be\label{phit}
 \delta^2_\eps\phi^I=-v^\nu \nabla_\nu\phi^I+[\phi^I, v^MA_M]-2 v^\nu \partial_\nu f \phi^I\,,
 \ee
 where the first two terms are  the negative of the Lie derivative and a gauge transformation. 
 
 Two supersymmetry transformations of the spinor $\Psi$  give
 \be\label{eq:2susyF1}
 \begin{split}
 \delta_\epsilon^2 \Psi\ = \ D_M\bkt{\eps \Gamma_N \Psi} \Gamma^{MN} \eps -2\,\p_\mu f\, \bkt{\eps \Gamma_I \Psi} \Gamma^{I \mu} \eps+ \delta_{\eps} K_m \nu^m
\end{split}
 \ee
 After using the CKSE in (\ref{KS1}) and the gamma matrix commutation relations, the first term in the above equation becomes
 \be
 \begin{split}
 D_M\bkt{\eps \Gamma_N \Psi} \Gamma^{MN} \eps=&\, d\,  \p_\mu f\, \eps\bkt{\eps \Gamma^\mu \Psi}-2 \p_\mu f\, \bkt{\eps \Gamma_\nu \Psi} \tilde{\Gamma}^\nu \Gamma^{\mu} \eps - \eps \bkt{\eps \slashed{D} \Psi} \\
& + \tilde{\Gamma}^M \Gamma^N\eps \bkt{\eps \Gamma_N D_M \Psi} + \p_\mu f\, \tilde{\Gamma}_\nu \Gamma_N \eps \bkt{ \eps \Gamma^N \tilde{\Gamma}^\mu \Gamma^\nu \Psi}  . 
\end{split}
 \ee
 The two terms on the last line can be simplified further using the triality condition in \cref{triality}, resulting in
  \be\label{eq:2susyT1}
 \begin{split}
 D_M\bkt{\eps \Gamma_N \Psi} \Gamma^{MN} \eps=&\, d\,  \p_\mu f\, \eps\bkt{\eps \Gamma^\mu \Psi}-2 \p_\mu f\, \bkt{\eps \Gamma_\nu \Psi} \tilde{\Gamma}^\nu \Gamma^{\mu} \eps - \eps \bkt{\eps \slashed{D} \Psi}
 -\frac12 v_N \tilde{\Gamma}^M \Gamma^N D_M \Psi\\ &
-\p_\mu f\, v^\mu \Psi 
+ \p_\mu f v_\nu \Gamma^{\nu\mu} \Psi
-\frac{d-2}{2} \p_\mu f v_N \tilde{\Gamma}^{N} \Gamma^\mu \Psi.
\end{split}
 \ee
 Using  the CKSE and triality, the second term in \cref{eq:2susyF1} becomes
 \be\label{eq:2susyT2}
  -2\,\p_\mu f\, \bkt{\eps \Gamma_I \Psi} \Gamma^{I \mu} \eps=\, \p_\mu f v_N \Gamma^{N \mu} \Psi -\p_\mu f\, v^\mu \Psi - 2\p_\mu\, f \bkt{\eps \Gamma_{\nu} \Psi} \tilde{\Gamma}^\mu \Gamma^\nu \eps. 
 \ee
 Using explicit form of $\delta_\eps K_m$ and the pure spinor properties \ref{psprop}, the third term in \cref{eq:2susyF1}  becomes
\be\label{eq:2susyT3}
\delta K_m \nu^m= \, \eps\bkt{\eps \slashed{D} \Psi} -\frac12 v_N \tilde{\Gamma}^N \Gamma^M D_M \Psi+ \frac12 \p_\mu \phi\, \eps \bkt{\eps \Gamma^\mu \Psi} -\frac14 \p_\mu \phi\, v_M \tilde{\Gamma}^M \Gamma^\mu \Psi.
\ee 
By combining \cref{eq:2susyT1,eq:2susyT2,eq:2susyT3} and using $\phi=-2\bkt{d-4} f$, we finally obtain the following. 
  \be \label{eq:twoPsi}
 \delta_\epsilon^2 \Psi\ = \ -v^\mu \nabla_\mu \Psi-\frac{1}{4}\nabla_{[\mu} v_{\nu]} \Gamma^{\mu\nu} \Psi + \sbkt{\Psi, v^M A_M}- 3 v^\mu\p_\mu f \Psi.
 \ee
 The first two terms constitute a Lie derivative of the spinor field. The third term is a gauge transformation and the last term is a Weyl transformation. 
 
 Similarly one can work out the action of two supersymmetry transformations on $K^m$.  Using properties of the pure spinors and the Bianchi identity $\sbkt{D_M,\sbkt{D_N,D_P}}=0$, we get
 \be 
 \delta_\epsilon^2 K^m\ =\ -v^\mu \nabla_\mu K^m + \sbkt{K_m, v^M A_M}- \bkt{\nu^{[m} \slashed{\nabla} \nu^{n]}} K_n\ -4 v^\mu\partial_\mu f K^m
 \ee
First two terms are Lie derivative and a gauge transformation respectively. The third term is an internal $SO(7)$ transformation which leaves the $\LL_{\text{aux}}$ invariant. The last term is a Weyl transformation. 
  Hence, we see that the non-trivial part of two supersymmetry variations acting on a field $\Phi$ is
 \be 
 \delta_\epsilon^2 \Phi \ =\ -\mathcal{L}_v -\Omega_\Phi \bkt{2 v^\mu\p_\mu f} \Phi\equiv- \delta_{v} \Phi.
 \ee
The  Weyl weights for different fields are
 \be
 \Omega_A=0,\quad \Omega_\phi=1, \quad \Omega_\Psi=\tfrac32,\quad \Omega_K=2.
 \ee

We next show that the action is invariant under the bosonic transformation $\delta_{v}$. 
To proceed, we note that $v^\mu$ is a
 conformal Killing vector which has constant components in stereographic coordinates.
\be
v^\mu = \eps \Gamma^{\mu} \eps\, = \frac{e^{\mu}{}_{\hat{\mu}}}{1+\beta^2 x^2} \eps_s \Gamma^{\hat{\mu}} \eps_s
\,
=
\delta^{\mu}{}_{\hat{\mu}} \eps_s \Gamma^{\hat{\mu}} \eps_s,
 \qquad
\nabla_\mu v_\nu =  4    \p_{[\mu} f \, v_{\nu ]} + 2 v^\rho \nabla_\rho f \, g_{\mu\nu}.
\ee
The transformation $\delta_v$ acts only on dynamical fields and leaves the background fields i.e., metric invariant.
\be
\delta_v g_{\mu\nu} = -\LL_v g_{\mu\nu} -\delta_{\text{Weyl}} g_{\mu\nu}\ =\ 0,
\ee
where
\be
\LL_v g_{\mu\nu}= 4v^\rho \nabla_\rho f \, g_{\mu\nu}, \qquad
\delta_{\text{Weyl}} \, g_{\mu\nu}\, = -4v^\rho \nabla_\rho f \, g_{\mu\nu}.
\ee
 The factor $\sqrt{g} e^{\phi}$ transforms as a scalar density of weight  $\frac{4}{d}$ under the action of Lie derivative, i.e.,
\be
\LL_v \sqrt{g}e^{\phi}= \sqrt{g}\left(v^\mu \p_\mu e^{\phi} + \frac{4}{d} e^{\phi} \p_\mu v^\mu\right).
\ee
Since $\p_\mu v^\mu =0$, this term essentially transforms as a scalar. All terms in the Lagrangian  multiplying the factor $e^\phi$ also transform as scalars  hence 
\be
\LL_{v} S \ =\ \int d^d x\, v^\mu \p_\mu \bkt{\sqrt{g} \LL}= \int d^d x\, \p_\mu \bkt{v^\mu \sqrt{g} \LL} =0.
\ee

We next look at the action of the Weyl transformations with respect to a parameter $\Omega$. It is easier to work with a finite version of the infinitesimal Weyl transformation appearing in $\delta_v$. Under a Weyl transformation $g_{\mu\nu}\to e^{2\Omega} g_{\mu\nu}$ the vector field does not transform. The rest of the fields transform as
 \be 
 \phi_I\to e^{-\Omega} \phi, \qquad \Psi\to e^{-\frac{3}{2}\Omega} \Psi, \qquad K^m\to e^{-2\Omega} K^m.
 \ee
 The Weyl transformations of different terms in the Lagrangian are
 \be \label{WT}
 \begin{split}
& g^{\mu\mu'} g^{\nu\nu'} F_{\mu\nu}F_{\mu'\nu'}\to e^{-4 \Omega}  g^{\mu\mu'} g^{\nu\nu'} F_{\mu\nu}F_{\mu'\nu'}, \\
&
g^{\mu\nu} \p_\mu\phi_I \p_\nu \phi^I\ \to e^{-4\Omega}\bkt{
g^{\mu\nu} \p_\mu\phi_I \p_\nu \phi^I\ + g^{\mu\nu} \p_\mu\Omega \p_\nu \Omega\ \phi_I \phi^I- 2 g^{\mu\nu} \phi^I \p_\mu \phi_I \p_\nu \Omega}, \\
& g^{\mu\nu} \sbkt{A_\mu,\phi_I}\sbkt{A_\nu,\phi^I}\to e^{-4\Omega} g^{\mu\nu} \sbkt{A_\mu,\phi_I}\sbkt{A_\nu,\phi^I}, \\
& \sbkt{\phi_I,\phi_J}\sbkt{\phi^I,\phi^J}\to e^{-4\Omega} \sbkt{\phi_I,\phi_J}\sbkt{\phi^I,\phi^J}, 
\\
&\Psi \slashed{D} \Psi\to e^{-4 \Omega}\bkt{\Psi \slashed{D} \Psi + \frac{d-4}{2} \p_\mu \Omega \Psi \Gamma^\mu \Psi }, \\ 
&K^m K_m\to e^{-4 \Omega} K^m K_m, \\ 
&\sqrt{g} e^{\phi}\to e^{4 \Omega} \sqrt{g} e^{\phi}.
 \end{split}
 \ee
 Note that the second term in the Weyl transformation for the fermion kinetic term is trivially  zero because  $\Gamma^\mu$ is a symmetric matrix and $\Psi$ is Grassmann odd.
 
 The Weyl transformation of the scalar mass terms  is more involved. To compute this we note that
 \be 
\frac{ d \Delta}{2 r^2} \phi_I \phi^I = \bkt{\frac{R}{d-1}+\frac{\bkt{d-4} }{4 d^2 } \p_\mu \log g \p^\mu \log g} \frac{\phi_I \phi^I}{2} ,
 \ee
 where $R$ is the Ricci scalar on the sphere. 
This  term transforms to
 \be 
 \begin{split}
 e^{-4 \Omega}\frac{ \phi_I\phi^I}{2}\Bigl(& \frac{R}{d-1} -2 \nabla^2 \Omega -2 \p_\mu\Omega \p^\mu \Omega \\
+& \frac{d-4}{d} \p_\mu \Omega \p^\mu \log g+ \frac{d-4}{4d^2}\p_\mu \log g\p^\mu \log g \Bigr) 
 \end{split}
 \ee
The third term cancels the second term coming from the scalar kinetic term in (\ref{WT}). Combining all scalar terms the total change in the action is given by 
  \be 
 \begin{split}
 \delta S\ = \int d^d x \sqrt{g} e^{\phi} \Bigl(&-\p_\mu \Omega \p^\mu \bkt{\phi_I \phi^I}   + \phi_I\phi^I\bkt{-\nabla^2 \Omega +\frac{d-4}{2d}\p^\mu \Omega \p_\mu\log g}
 \Bigr)
  \end{split}
  \ee
  Integrating the first term  by parts and using $\phi=\frac{4-d}{2d}\log g$ we see that total variation of the action is zero
  \be
  \delta S =0.
  \ee
  Hence we have shown that the supersymmetry algebra closes off-shell for a particular supersymmetry parameter $\eps$. 
This conclusion holds true more generally on-shell when one considers the anti-commutator of two supersymmetry variations w.r.t parameters $\eps_1$ and $\eps_2$. In that case  $v^\mu=\eps_1 \Gamma^\mu \eps_2$.

\section{Superalgebra considerations}
\label{sec:salgebra}

In this section we examine the symmetries of the theories constructed above in detail and identify the underlying superalgebras. We will show that the symmetry algebra of these theories 
is isomorphic to the maximal Poincar\'e superalgebra $siso(d)$.

\subsection{The Poincar\'e superalgebra}
Let us start by reviewing some basic facts about $siso(d)$. 
This is not a semisimple superalgebra as the even part of the algebra has an abelian ideal which contains  translations \cite{Nahm:1977tg}.
The abelian ideal also contains a certain number of $p$-form charges which commute with translations but transform as tensors of $so(d)$. 
They have interpretation  in terms of $p$-brane charges \cite{0305-4470-15-12-028,Townsend:1995gp} (see \cite{Yokoyama:2015yga} for some explicit computations  in SYM). 
 The bosonic part of $siso(d)$ may also contain an automorphism algebra. Part of the automorphism algebra under which the fundamental fields transform non-trivially is called the $R$-symmetry algebra. The odd part of the algebra contains sixteen supersymmetry generators which transform under the spinor representation of $SO\bkt{d}$.
 
  Schematically the (anti-)commutation relations for $siso(d)$ take the following form (ignoring the $p$-form central charges):
\be\label{eq:schsiso}
\begin{split}
\sbkt{M,M}\sim M,\quad \sbkt{M,P}\sim M,\quad \sbkt{P,P}\sim 0,  \\
\{ Q,Q\}\sim P+Z,\quad \sbkt{M,Q}\sim Q, \quad \sbkt{P,Q}\sim 0,  \\
\sbkt{R,R}\sim R\quad\sbkt{M,R}\sim 0,\sbkt{Q,R}\sim Q, \\
\sbkt{R,Z}\sim Z,\quad \sbkt{M,Z} \sim \sbkt{P,Z}\sim\sbkt{Z,Z}\sim 0.
\end{split}
\ee
where the first line is Poincar\'e algebra, the second line is the super translation algebra, the third line is the automorphism algebra and the last line contains the  transformation of the central charges
under the rest of the bosonic generators. Strictly speaking, the above form of the algebra is only true when acting on gauge invariant operators of the theory. In general, the action of  $\{ Q,\ Q\}$ on the fundamental fields of the theory may involve a field-dependant gauge transformation~\cite{deWit:1975veh,Ferrara:1976iq}. Upon spontaneous symmetry breaking, the field-dependant gauge transformations give way to zero-form central charges~\cite{Fayet:1978ig}. Henceforth, we  set all central charges to zero. 
More details about the algebra underlying 
maximally supersymmetric Yang-Mills in Minkowski signature can be found in \cite{Seiberg:1997ax}.

\subsection{Symmetry algebra for theories with non-constant coupling}
Let us now turn to the symmetries of the theories constructed in this paper. We will demonstrate that this symmetry algebra is isomorphic to $siso(d)$.

\subsubsection{Bosonic symmetries}
\subsubsection*{\underline{SO(d)}}
The isometry group of $\bS^d$ can be realized by the standard embedding in $\bR^{d+1}$. Let $X^A, A=1,2,\cdots d+1$ be coordinates on $\bR^{d+1}$. The isometry group of the sphere is then generated by 
\be
M_{AB}= X_A \p_B-X_B \p_A,
\ee
subject to the embedding $\sum_A X^A X^A=r^2$. The stereographic coordinates $x^\mu, \mu=1,\cdots $ are given by
\be
x^\mu= 2\frac{X^\mu}{r-X^{d+1}}, \quad X^\mu= \frac{ x^\mu}{1+\beta^2 x^2},\qquad X^{d+1}= r\frac{\beta^2x^2-1}{1+\beta^2 x^2}.
\ee
In stereographic coordinates, the generators of the isometry group take the following form
\be
M_{\mu\nu} =x_\mu \p_\nu-x_\nu\p_\mu,\qquad 
\hat{M}_{\mu,d+1}=  2 \beta^2 x_\mu x_\rho \p_\rho - \bkt{\beta^2 x^2-1} \p_\mu.
\ee
The generators $\hat{M}_{\mu,d+1}$ do not leave the coupling invariant (except for $d=4$) and hence 
do not 
generate a bosonic symmetry of the theory. 
 $M_{\mu\nu}$  generates the $SO(d)$ subgroup of the isometry group of $\bS^{d}$ under which the Lagrangian is invariant. In the expression for $M_{\mu\nu}$, the index for the coordinate is lowered with the flat metric, i.e., $x_\mu= x^\mu$. The theory on $\bS^{d}$ also has an $R$-symmetry $SO\bkt{1,9-d}$, which acts on the scalars.

\subsubsection*{\underline{$\bR^d$}}

In our construction there is a bosonic symmetry generated by the vector field $v=v^\mu\p_\mu$, with constant components $v^\mu$ in stereographic coordinates. The transformation of the fields under $v$  
takes the following form
\be
\delta_v\ =\ \LL_v+  2 \Omega v^\mu\p_\mu f,
\ee
where $\Omega$ is the Weyl weight of the field. 
 The Weyl transformations for different $v$'s commute trivially. Using the fact that $v$'s have constant components in stereographic coordinates one can explicitly show that
 \be
 \sbkt{\delta_{v_1}, \delta_{v_2}} =0.
 \ee
 Hence, the transformations generated by $v$ form an abelian group isomorphic to $d$-dimensional translations. 
  
 Under the $SO\bkt{d}$ group, the field transformations are given by Lie derivatives w.r.t the vector fields $M_{\mu\nu}$.
 \be
 \delta_{M}= \LL_{M}.
 \ee
 When acting on a field $\Phi$ the commutator of $\delta_M$ and $\delta_v$ is given by
 \be
 \sbkt{\delta_{M},\delta_v} \Phi = \sbkt{\LL_{M},\LL_v}\Phi+ 2 \Omega_\Phi \LL_M \bkt{v^\rho \p_\rho f} \Phi.
 \ee
 Using $\p_\rho f=-\frac{\beta^2 x^\rho}{1+\beta^2 x^2}$ we get
 \be
 \LL_M\bkt{v^\rho \p_\rho f} = v^\nu \p_\mu f-v^\mu \p_\nu f.
 \ee
 Also by using 
 \be
\sbkt{M_{\mu\nu},v}= v^\nu\p_\mu-v^\mu \p_\nu,
 \ee
 we see that
 \be
  \sbkt{\delta_{M},\delta_v} \Phi= \delta_{\sbkt{M,v}} \Phi.
 \ee
 Hence we see that the group of transformations generated by $M$ and $v$ is isomorphic to the Poincar\'e group.
 
 \subsubsection{Fermionic symmetries}
 Let us now analyze the fermionic symmetries. Supersymmetry parameters transform under the spinor representation of $SO(d)$. 
 We have already demonstrated that the anti-commutator of two supersymmetry transformations generates a symmetry transformation w.r.t $v$ (up to gauge transformations). Let us now work out the commutator of a supersymmetry transformation and a transformation generated by $v$. For bosonic fields $A_{M}$, acting with $ \delta_\eps$ followed by $\delta_v$ we get:
 \be
 \delta_v \delta_\eps A_M= \bkt{\eps \Gamma_M \LL_v \Psi} +3 v^\rho\p_\rho f \bkt{\eps \Gamma_M \Psi}.
 \ee
 If we do the two transformations in opposite order we get the following for the vector field,
 \be
 \delta_\eps \delta_v A_\mu = \LL_v\bkt{\eps \Gamma_\mu \Psi}= \bkt{\LL_v \eps \Gamma_\mu \Psi}+\eps\bkt{\LL_v \Gamma_\mu} \Psi +\bkt{\eps \Gamma_\mu \LL_v \Psi}
 \ee
 Using the conformal Killing spinor equation and the definition of the spinorial Lie-derivative we have
 \be
  \bkt{\LL_v \eps \Gamma_\mu \Psi}= v^\rho\p_\rho f \bkt{\eps \Gamma_\mu \Psi},
 \ee
while using $\Gamma_\mu= e_\mu{}^{\hat{\mu}} \Gamma_{\hat{\mu}}$  
we get
 \be
 \eps\bkt{\LL_v \Gamma_\mu} \Psi= 2v^\rho\p_\rho f \bkt{\eps \Gamma_\mu \Psi}.
 \ee
Combining these results then gives
 \be
 \delta_\eps \delta_v A_\mu =  \bkt{\eps \Gamma_\mu \LL_v \Psi} +3 v^\rho\p_\rho f \bkt{\eps \Gamma_\mu \Psi}.
 \ee
 Similarly, acting on the scalars $A_I=\phi_I$, we have
 \be
  \delta_\eps \delta_v \phi_I =  \LL_v\bkt{\eps \Gamma_I \Psi}+ 2 v^\rho\p_\rho f \bkt{\eps \Gamma_I \Psi} =   \bkt{\eps \Gamma_I \LL_v \Psi} +3 v^\rho\p_\rho f \bkt{\eps \Gamma_I \Psi}.
 \ee
Hence,
 \be
 \sbkt{\delta_\eps , \delta_v} A_M =0.
 \ee
 
 Let us now compute this commutator acting on the fermion. 
 \be
 \begin{split}
 \delta_v\delta_\eps \Psi&= \delta_v \bkt{\tfrac12 F_{\mu\nu}\Gamma^{\mu\nu} + \bkt{D_\mu \phi_I + 2 \p_\mu f \phi_I} \Gamma^{\mu I}+ \tfrac12 F_{IJ}\Gamma^{IJ}} \eps \\
 &=
 \LL_{v} F_{\mu\nu}\, \Gamma^{\mu\nu} \eps + \LL_v\bkt{D_\mu\phi_I + 2 \p_\mu f \phi_I} \Gamma^{\mu I }\eps+\tfrac12 \LL_v F_{IJ}\, \Gamma^{IJ} \eps \\
 &\quad + 2v^\rho\p_\rho f \bkt{D_\mu \phi_I +2 \p_\mu f }\Gamma^{\mu I} \eps + 2 v^\rho\p_\rho f F_{IJ} \Gamma^{IJ} \eps. 
 \end{split}
 \ee
 Acting with the transformations in the opposite order, we get
 \be
 \delta_\eps \delta_v \Psi = \LL_v \delta_\eps \Psi + 3v^\rho\p_\rho f \delta_\eps \Psi.
 \ee
 A simple computation gives the following for the first term on the r.h.s
 \be
 \begin{split}
 \LL_v \delta_\eps \Psi&= \tfrac12 \LL_v F_{\mu\nu} \Gamma^{\mu\nu} \eps -\tfrac32 v^\rho \p_\rho f F_{\mu\nu} \Gamma^{\mu\nu} \eps \\
 &\quad +  \LL_v\bkt{D_\mu\phi_I + 2 \p_\mu f \phi_I} \Gamma^{\mu I }\eps-v^\rho\p_\rho f \bkt{D_\mu\phi_I + 2 \p_\mu f \phi_I} \Gamma^{\mu I }\eps\\
 &\quad + \tfrac12 \LL_v F_{IJ}\, \Gamma^{IJ} \eps +\tfrac12 v^\rho \p_\rho F_{IJ}\, \Gamma^{IJ} \eps.
 \end{split}
 \ee
 In evaluating the above we have used properties of the Lie-derivative, the explicit forms of the vielbeins, and the CKSE. From these one finds that
 \be
 \sbkt{\delta_\eps ,\delta_v} \Psi =0,
 \ee
establishing the isomorphism. 
\section{Comments on  localizing  the path integral} \label{sec:localization}
In this section we give a preliminary analysis of path integral  localization  for the theories constructed in this paper. We start with a straightforward generalization of \cite{Pestun:2007rz} for general $d$ that gives a localization locus where
scalar fields have a non-constant profile.  This profile  leads to a divergent action for  $d>2$.  However, for $d=2$ the localized action is finite and the partition function reduces to an integral over constant matrices.  If $d<2$ then the localized action is zero.

We then specialize to $d=4$ where we choose a different localization Lagrangian.  This procedure  is akin to the Higgs branch localization of \cite{Benini:2012ui} in the case of two-dimensional $(2,2)$ theories.

We start by choosing our localization Lagrangian to be $
Q\cdot V$, where
\be
V\equiv \int d^d x \sqrt{g}\Tr \Psi \overline{\delta_{\epsilon} \Psi},
\ee
and $  \overline{\delta_\epsilon\Psi}$ is \emph{defined} as
\be
  \overline{\delta_\epsilon\Psi} = 
        \frac{1}{2}\tilde{\Gamma}^{MN}F_{MN}\Gamma^0\epsilon
        +\frac{\alpha_I}{2}\,\tilde{\Gamma}^{\mu I}\phi_I\Gamma^0\nabla_\mu\epsilon
        -K^m\Gamma^0 \nu_m.
        \ee
 We can split the localization Lagrangian into two parts which receive contributions from bosonic and fermionic degrees of freedom,   
\be 
\begin{split}
Q\cdot V= \int d^d x \sqrt{g} \Tr \delta_\eps \Psi \delta_\eps \overline{\Psi} -\int d^d x \sqrt{g} \Tr \Psi \delta_\eps^2 \overline{\Psi}   \equiv\int d^d x \sqrt{g}  \bkt{\LL^{\text{b}}+ \LL^{\text{f}}}\,.
\end{split}
\ee
We then compute the bosonic part explicitly. The computation is analogous to the one  in \cite{Minahan:2015jta}, resulting in
\be \label{eq:LocB}
\begin{split}
\mathcal{L}^{b}\ =\Tr\Bigl(  &\frac{v^0}{2} F^{MN}F_{MN} -\frac{1}{4} F^{MN}F^{PQ} \bkt{\eps \Gamma^{MNPQ0} \eps} \\
& -2 \phi_I F_{MN} \p_\mu f \bkt{\epsilon \Gamma^{\mu I M N 0}\eps} \\
&-4 \phi_I F^{I \mu} \p_\mu f v^0 + 4 v^0 \phi_I \phi^I \p_\mu f\p^\mu f - K^m K_m v^0\Bigr).
\end{split}
\ee
We next write the  scalar field terms on the last line of equation (\ref{eq:LocB}) as 
\be 
\begin{split}
-4 \phi_I F^{I \mu} \p_\mu f v^0 + 4 v^0 \phi_I \phi^I \bkt{\p_\mu f}^2 v^0&= 4 v^0 \bkt{\phi_I \p_\mu f +\frac{1}{2} D_\mu \phi_I} \bkt{\phi^I \p^\mu f +\frac{1}{2} D^\mu \phi^I}
\\
& -v^0 F^{\mu I} F_{\mu I},
\end{split}
\ee
where the last term  cancels against the scalar kinetic term  in $\frac{1}{2}v^0 F_{MN}F^{MN}$. Hence, the bosonic part of the localization Lagrangian takes the  form
\be 
\begin{split}
\mathcal{L}^{\text{b}}  = 
\Tr\Bigl( & \frac{v^0}{2}F^{\mu\nu} F_{\mu\nu} -\frac{1}{4} F_{\mu\nu}F_{\rho\sigma} \bkt{\eps \Gamma^{\mu\nu\rho\sigma 0} \eps} +\frac{v^0}{2} F^{IJ}F_{IJ} -\frac{1}{4} F_{IJ}F_{KL} \bkt{\eps \Gamma^{IJKL 0} \eps} \\
&-\frac{1}{2} F_{\mu I}F_{MN} \bkt{\eps \Gamma^{\mu I M N 0} \eps} -2 \phi_I F_{MN} \p_\mu f \bkt{\epsilon \Gamma^{\mu I M N 0}\eps} \\
&  + 4 v^0 \bkt{\phi_I \p_\mu f +\frac{1}{2} D_\mu \phi_I} \bkt{\phi^I \p^\mu f +\frac{1}{2} D^\mu \phi^I} -K^m K_m v^0
\Bigr).
\end{split}
\ee

To simplify things, let us limit ourselves to the zero instanton sector, where $A_\mu=0$. 
In this case the localization Lagrangian reduces to
\be \label{bosonicfp}
\begin{split}
\mathcal{L}^{\text{b}}  = 
&\Tr\Big(- \nabla_\mu \phi_I \nabla_\nu \phi_J\bkt{\eps \Gamma^{\mu I \nu J 0} \eps} -4 \phi_I \nabla_\nu \phi_J \p_\mu f \bkt{\epsilon \Gamma^{\mu I \nu J 0}\eps} \\
&  + 4 v^0 \bkt{\phi_I \p_\mu f +\frac{1}{2} \nabla_\mu \phi_I} \bkt{\phi^I \p^\mu f +\frac{1}{2} \nabla^\mu \phi^I}\\
&+\frac{v^0}{2} F^{IJ}F_{IJ} -\frac{1}{4} F_{IJ}F_{KL} \bkt{\eps \Gamma^{IJKL 0} \eps}-K^m K_m v^0\Big)
.
\end{split}
\ee
After a field redefinition $\phi_I = e^{-2 f} \phi'_I$, the quadratic terms become
\be \label{eq:lbfr}
\begin{split}
\mathcal{L}^{\text{b}}  &= \Tr\Big(e^{-4 f}\nabla_\mu \phi'_I \Bigl( v^0 \nabla^\mu \phi'^I
-  \nabla_\nu \phi'_J\bkt{\eps \Gamma^{\mu I \nu J 0} \eps}
\Bigr)-K^m K_m v^0\Big).
\end{split}
\ee
This term is positive definite after Wick rotating  $\phi^0$ and $K_m$.   It is  minimized to zero if 
\be
\phi_I'= \beta\,\sigma_{I}\,,\qquad K_m=0, 
\ee
where $\sigma_I$ are constant elements of the Lie algebra.   The quartic terms in (\ref{bosonicfp}) force the $\sigma_I$ to belong to the Cartan subalgebra.  

Since the auxiliary fields are set to zero, the localization locus is on-shell.  In fact, the locus is the usual on-shell solution to the equations of motion in flat space after a conformal transformation from the sphere.  As such, unlike the usual localization prescription, it is possible to have all scalars  non-zero at the locus.

After substituting the locus  into the Lagrangian in (\ref{LL1}) one finds that the action is 
\bea\label{actionloc}
S&=&\frac{1}{g_{YM}^2}\int d^dx\frac{1}{(1+\beta^2x^2)^2}\Tr\Big((1+\beta^2x^2)\partial_\mu\phi'_I\partial^\mu\phi'^I+2\beta^2(1+\beta^2x^2)x^\mu\partial_\mu(\phi'_I\phi'^I)\nn\\
&&\qquad\qquad\qquad\qquad\qquad\qquad\qquad\qquad+2\beta^2(d+(d-2)\beta^2x^2)\phi'_I\phi'^I\Big)\nn\\
&=&\frac{1}{g_{YM}^2}\int d^dx\frac{2\beta^4(d+(d-2)\beta^2x^2)}{(1+\beta^2x^2)^2}\Tr(\sigma_I\sigma^I)
\eea
If $d>2$ then this integral diverges.  We emphasize that the behavior of the coupling at the poles is \emph{not} responsible for this divergence. In fact, one finds a divergent action even for  $d=4$ where the coupling is constant. 

If  $d<2$ then (\ref{actionloc}) is zero.  
However, if $d=2$ then it is finite and nonzero, given by
\be\label{actionlocd2}
S=\frac{4\pi\beta^2}{g_{YM}^2}\Tr \s_I\s^I\,.
\ee
At first the result in (\ref{actionlocd2}) might seem puzzling, since we are finding a nonzero answer even though the equations of motion are satisfied, which is what happens for $d<2$.  This is because the equations of motion are obtained by an integration by parts, which could lead to boundary terms.  At the north pole where the coupling is finite the boundary term is zero.  But at the south pole where the coupling is divergent for $d=2$, there is a contribution, which then gives (\ref{actionlocd2}).  From the flat space point of view, the nontrivial contribution from the locus can be attributed to the inclusion of the point at infinity.
For $d=2$ there are no instanton contributions so we expect no other localization fixed points.  It is now just a question of computing the determinant factors, which we do not attempt in this paper.

We now turn to a different localization procedure.
We will still localize with respect to the same supercharge as above but a modified localization term. 
It is reasonable to assume that the fields at the localization locus only depend on the polar angle.
 We note that the second term in \cref{eq:lbfr}  vanishes for such configurations. We now modify the localization Lagrangian by adding the term
 \be\label{QVF}
Q\cdot V'=\delta\Tr \bkt{ \overline{ \Psi} \eps F\bkt{\phi}},
 \ee
 where $F\bkt{\phi}$ is a functional which depends on the scalar fields. We will make a specific choice for $F\bkt{\phi}$, which is akin to the Higgs branch localization of \cite{Benini:2012ui}. For vanishing gauge fields the bosonic part of (\ref{QVF}) is
 \be
 Q\cdot V'|_{\text{bos}}=\Tr(v^\mu F\bkt{\phi}\bkt{\p_\mu+2\p_\mu f} \phi^0 )
 \ee
where 
 \be
 F\bkt{\phi}\equiv -2\frac{\beta^2 x^2}{v^\mu x^\mu} \phi^0,
 \ee
 The denominator in the above term is proportional to the component of the vector fields $v^\mu$ along the polar angle. With this choice the full localization action becomes 
 \be
 \int \sqrt{g}\, \bkt{ Q\cdot V|_{\text{bos}}+ Q\cdot V'|_{\text{bos}}} \ =
 \int \sqrt{g} \tfrac{1}{1+\beta^2 x^2}\Tr( \phi^0 \nabla^2 \phi^0) + 2 \beta^2 \int \sqrt{g} \bkt{x^\mu-\frac{ x^2 v^\mu}{v^\rho x^\rho}} \Tr(\phi^0\bkt{ \p_\mu+2\p_\mu f}\phi^0).
 \ee
 The last term  vanishes on configurations which depend only on the polar angle.
 Since the Laplacian on the sphere is a negative-semi-definite operator, we conclude that for imaginary $\phi^0$ the above localization Lagrangian is positive-semi-definite. The zero eigenvalue corresponds to a constant value of the scalar field.

\section{Conclusions and outlook}\label{sec:conc}
In this paper we have constructed a class of  maximally
supersymmetric gauge theories on spheres, where the gauge coupling explicitly depends on the sphere's polar angle.  Our constructions can be viewed as natural IR regularizations of flat space theories. We have further shown that the symmetry algebra of these theories is isomorphic to a corresponding Poincar\'e superalgebra.  We have also presented a preliminary localization analysis for these theories.

There are various avenues for further research.  
One   
issue is the computation of the partition functions 
using localization.  
The perturbative partition function for gauge theories with eight and sixteen supercharges on general $\bS^d$ is known \cite{Minahan:2015any,Minahan:2017wkz,Gorantis:2017vzz}. 
It would be   
interesting to compute the partition function for theories with position dependent coupling and compare it to   
maximally supersymmetric theories with constant coupling.
Here we would expect the partition functions to depend on the dimensionless parameter $g_{\text{YM}}^2/r^{d-4}$ where $g_{\text{YM}}^2$ is the coupling at the north pole. 
Our 
analysis indicates that unless  $d=2$, the best candidate for a suitable localization term is not the standard choice, i.e., $\delta \Psi\delta \overline{\Psi}$.
It will be instructive to reproduce the known localization results for $\NN=4$ SYM by localizing w.r.t a Poincar\'e supercharge.
 Moreover, it is also reasonable to expect that the position dependence of the coupling  appears --- in some form or another --- in the quadratic fluctuations about the localization locus. 
We expect that an approach based on index-theorem would be the most feasible to compute one-loop determinants.

Another interesting direction is to make contact with holographic duals of maximal SYM. Recently holographic duals of maximal SYM on spheres with constant coupling were constructed in \cite{Bobev:2018ugk}. 
It is natural to extend the analysis of \cite{Bobev:2018ugk} and find the holographic duals of theories constructed in this paper. 
We anticipate that holographic duals of these theories are supergravity solutions with non-constant profiles for the dilaton along the sphere directions.
It will be interesting to work out in detail how the position dependence of the coupling manifests itself in the holographic duals.

 The five and six dimensional versions of our construction are of particular interest since these theories have interesting UV completions. The 6D MSYM theory is related to  (1,1) little string  theory~\cite{Seiberg:1997zk,Aharony:1998ub}.   
By compactifying on a circle, 6D (2,0)  theory reduces to 5D MSYM. In this way, 5D theory can be used to study properties of 6D (2,0) theory. For example, in \cite{Kim:2012ava,Kallen:2012zn} the $N^3$-behavior of the free energy was derived using the MSYM on $\bS^5$ with constant coupling. 
  There is, however, a mismatch between the coefficient of the $N^3$ term on two sides. Perhaps the 5D theory constructed here can 
  help resolve this mismatch.

  \section{Acknowledgements}
 We  thank Takuya Okuda for comments on an earlier draft. The research of  J.A.M.  is supported in part by
Vetenskapsr{\aa}det under grant \#2016-03503 and by the Knut and Alice Wallenberg Foundation under grant Dnr KAW 2015.0083.
The work of U.N is supported by the U.S. Department of Energy, Office of Science, Office of High Energy Physics of U.S. Department of Energy under grant Contract Number  DE-SC0012567
and the fellowship by the Knut and Alice Wallenberg Foundation, Stockholm Sweden. J.A.M. thanks the CTP at MIT  for kind
hospitality during the course of this work.

\appendix\section{Clifford algebra conventions}\label{conv}

We use 10-dimensional Majorana-Weyl spinors $\eps_\al$, $\Psi_\al$ {\it etc.}.  Spinors in the other representation are written with a tilde, $\tilde\eps^\al$, {\it etc.}  The 10-dimensional $\Gamma$-matrices are chosen to be real and symmetric, 
\be
\Gamma^{M\al\beta}=\Gamma^{M\beta\al}\qquad \tilde\Gamma^M_{\al\beta}=\tilde\Gamma^M_{\beta\al}\,.
\ee
Products of $\Gamma$-matrices are given by
\be
\begin{split}
&\Gamma^{MN}\equiv \tilde\Gamma^{[M}\Gamma^{N]}\qquad \tilde\Gamma^{MN}\equiv \Gamma^{[M}\tilde\Gamma^{N]}  \\
&\Gamma^{MNP}\equiv\Gamma^{[M}\tilde\Gamma^{N}\Gamma^{P]}\qquad\tilde\Gamma^{MNP}\equiv\tilde\Gamma^{[M}\Gamma^{N}\tilde\Gamma^{P]}\,,\ \mbox{\it etc.}
\end{split}
\ee
We also have that $\Gamma^{MNP\al\beta}=-\Gamma^{MNP\beta\al}$, hence
\be
\eps\Gamma^{MNP}\eps=0\,
\ee
for any bosonic spinor $\eps$.

A very useful relation is the triality condition,
\be\label{triality}
\Gamma^M_{\al\beta}\Gamma_{M\gamma\delta}+\Gamma^M_{\beta\delta}\Gamma_{M\gamma\al}
+\Gamma^M_{\delta\al}\Gamma_{M\gamma\beta}=0\,.
\ee
We can use this to show that
\be
\eps\Gamma^M\eps\,\eps \Gamma_M\chi=0\,,
\ee
where $\chi$ is any spinor.  It immediately follows that $v^Mv_M=0$, where $v^M$ is the vector field
\be
v^M\equiv \eps\Gamma^M\eps\,.
\ee

We also use a set of pure-spinors, $\nu_m$ which satisfy the properties
\be\label{psprop}
\begin{split}
\nu_m\Gamma^M\eps=& 0  \\
\nu_m\Gamma^M\nu_n=&\delta_{mn}v^m  \\
\nu^m_\al\nu^m_\beta+\eps_\al\eps_\beta=&\frac12 v^M\tilde\Gamma_{M\al\beta}\,.
\end{split}
\ee
They are invariant under an internal $SO(7)$ symmetry, which can be enlarged to $SO(8)$ by including $\eps$.

\section{SUSY variation of Lagrangian on $\bS^d$}
\label{app:delLgenD}
Under supersymmetry transformations of \cref{susysp} the Lagrangian in \cref{LL1} changes as follows:
\be
\begin{split}
\delta\LL=&\frac{e^\phi}{g_{\text{YM}}^2}\Big(2F^{MN}D_M(\eps\Gamma_N\Psi)+\sfrac{1}{2}\left(F^{MN}\eps\Gamma_{MN}+\sfrac{4}{d}(\nabla^\mu\eps)\Gamma_{\mu I}\phi^I\right)\slashed{D}\Psi \\
&\qquad+\sfrac{1}{2}\Psi\slashed{D}\left(F^{MN}\eps\Gamma_{MN}+\sfrac{4}{d}(\nabla^\mu\eps)\Gamma_{\mu I}\phi^I\right)
+4d\beta^2\Delta\,\phi^i\eps\Gamma_I\Psi\Big)
\end{split}
\ee
We do an integration by part on the first term in the second lines and move the fermion field to the right to get
\be
\begin{split}
\delta \LL
=&\frac{e^\phi}{g_{\text{YM}}^2}\Big(2F^{\mu N}\tilde\eps\Gamma_\mu\Gamma_N\Psi+2F^{MN}\eps\Gamma_N D_M\Psi
+
F^{MN}\eps\Gamma_{MN}\slashed{D}\Psi
+
4\tilde\eps \Gamma_{ I}\phi^I\slashed{D}\Psi
+4d\beta^2\Delta\, \phi^I\eps\Gamma_I\Psi
  \\
&\qquad+\frac{1}{2}\partial_\nu \phi \left(F^{MN}\eps\Gamma_{MN}\Gamma^\nu\Psi+4\phi^I\tilde\eps\Gamma_I\Gamma^\nu\Psi\right)\Big)\\
=&\frac{e^\phi}{g_{\text{YM}}^2}\Big(2F^{\mu N}\tilde\eps\Gamma_\mu\Gamma_N\Psi+F^{MN}\eps\Gamma_{MNL}D^L\Psi
+
4\phi^I\tilde\eps\Gamma_I\slashed{D}\Psi
+
4d\beta^2\Delta\,\phi^I\eps\Gamma_I\Psi  \\
&\qquad \qquad
+\frac{1}{2}\p_\nu \phi\left(F^{MN}\eps\Gamma_{MN}\Gamma^\nu\Psi+4\phi^I\tilde\eps\Gamma_I\Gamma^\nu\Psi\right)\Big)\,.
 \end{split}
\ee
Upto total derivatives, the second term can be written as
\be
e^\phi F^{MN}\eps\Gamma_{MNL}D^L\Psi=-e^\phi\partial_\nu \phi F^{MN}\eps{\Gamma_{MN}}^\nu\Psi-e^\phi F^{MN}\tilde\eps \Gamma^\nu\Gamma_{MN\nu}\Psi.
\ee
This can be further simplified by using 
\be
F^{MN}\tilde\eps \Gamma^\nu\Gamma_{MN\nu}\Psi=(d-2)F^{\mu\nu}\tilde\eps\Gamma_{\mu\nu}\Psi+2(d-1)F^{\mu I}\tilde\eps\Gamma_{\mu I}\Psi+d F^{IJ}\tilde\eps\Gamma_{IJ}\Psi\,.
\ee
Similarly up to total derivatives, we have
and
\be
\begin{split}
4e^\phi \phi^I\tilde\eps\Gamma_I\slashed{D}\Psi
&=4e^\phi \phi^I\tilde\eps\Gamma_I\Gamma^\mu D_\mu \Psi+4e^\phi \phi^I\tilde\eps\Gamma_I\Gamma_J \sbkt{\phi_J, \Psi}
\\
&= 
-4 e^\phi \p_\mu \phi \phi^I \tilde{\eps} \Gamma_{I\mu} \Psi+4 e^\phi F^{\mu I} \tilde{\eps} \Gamma_{\mu I} \Psi
-4e^\phi \phi^I \bkt{\nabla_\mu \tilde{\eps}} \Gamma_I \Gamma^\mu \Psi +4 e^\phi F^{IJ} \tilde{\eps} \Gamma_{IJ} \Psi. 
\end{split}
\ee
In the second equality, the last term arises because $\Tr\bkt{\phi_I \sbkt{\phi_J,\Psi}}=\Tr\bkt{\sbkt{\phi_I,\phi_J} \Psi}$.
Combining these terms, we get
\be\label{eq:delLgenD}
\begin{split}
\delta\LL=&\frac{e^\phi}{g_{\text{YM}}^2}\Big(-(d-4)F^{MN}\tilde\eps\Gamma_{MN}\Psi-4\phi^I(\nabla^\mu\tilde\eps)\Gamma_{I\mu}\Psi
+4d\beta^2\Delta\,\phi^I\eps\Gamma_I\Psi
 \\
&\qquad\frac{1}{2}\p_\nu \phi \left(F^{MN}\eps\Gamma^{MN}\Gamma^{\nu}\Psi-2F^{MN}\eps\Gamma^{MN\nu}\Psi-4\phi^I\tilde\eps\Gamma^I\Gamma^\nu\Psi\right) 
\Big).
\end{split}
\ee

\bibliographystyle{JHEP} 
\bibliography{refs}  
 
\end{document}